\documentclass[12pt]{article}

\usepackage{amsfonts}
\usepackage[centertags]{amsmath}
\usepackage{amssymb}
\usepackage{cite}
\usepackage{psfrag}
\usepackage{graphicx}
\usepackage{bbm,bm}
\usepackage{epsfig, multicol}

\allowdisplaybreaks
\begin{document}

\topmargin 0pt
\oddsidemargin 0mm
\def\be{\begin{equation}}
\def\ee{\end{equation}}
\def\bea{\begin{eqnarray}}
\def\eea{\end{eqnarray}}
\def\ba{\begin{array}}
\def\ea{\end{array}}
\def\ben{\begin{enumerate}}
\def\een{\end{enumerate}}
\def\nab{\bigtriangledown}
\def\tpi{\tilde\Phi}
\def\nnu{\nonumber}
\newcommand{\eqn}[1]{(\ref{#1})}

\newcommand{\half}{{\frac{1}{2}}}
\newcommand{\vs}[1]{\vspace{#1 mm}}
\newcommand{\dsl}{\pa \kern-0.5em /} 
\def\a{\alpha}
\def\b{\beta}
\def\g{\gamma}\def\G{\Gamma}
\def\d{\delta}\def\D{\Delta}
\def\ep{\epsilon}
\def\et{\eta}
\def\z{\zeta}
\def\t{\theta}\def\T{\Theta}
\def\l{\lambda}\def\L{\Lambda}
\def\m{\mu}
\def\f{\phi}\def\F{\Phi}
\def\n{\nu}
\def\p{\psi}\def\P{\Psi}
\def\r{\rho}
\def\s{\sigma}\def\S{\Sigma}
\def\ta{\tau}
\def\x{\chi}
\def\o{\omega}\def\O{\Omega}
\def\k{\kappa}
\def\pa {\partial}
\def\ov{\over}
\def\nn{\nonumber\\}
\def\ud{\underline}
\begin{flushright}
%
\end{flushright}
\begin{center}
{\large{\bf Space-like D$p$ branes: accelerating cosmologies\\ versus      
conformally de Sitter space-time}}

\vs{10}

{Kuntal Nayek\footnote{E-mail: kuntal.nayek@saha.ac.in} and Shibaji Roy\footnote{E-mail: shibaji.roy@saha.ac.in}}

\vs{4}

{\it Saha Institute of Nuclear Physics\\
1/AF Bidhannagar, Calcutta 700064, India\\}

\end{center}

\vs{15}

\begin{abstract}
We consider the space-like D$p$ brane solutions of type II string theories having isometries ISO$(p+1)$
$\times$ SO$(8-p,1)$. These are asymptotically flat solutions or in other words, the metrics become flat at the 
time scale $\tau \gg \tau_0$. On the other hand, when $\tau \sim \tau_0$, we get $(p+1)+1$ dimensional flat FLRW
metrics upon compactification on a $(8-p)$ dimensional hyperbolic space with time dependent radii. We show
that the resultant $(p+1)+1$ dimensional metrics describe transient accelerating cosmologies for all
$p$ from 1 to 6, i.e., from $(2+1)$ to $(7+1)$ space-time dimensions. We show how the accelerating phase changes with
the interplay of the various parameters characterizing the solutions in $(3+1)$ dimensions. Finally,
for $\tau \ll \tau_0$, after compactification on $(8-p)$ dimensional hyperbolic space, the resultant metrics
are shown to take the form of  $(p+1)+1$ dimensional de Sitter spaces upto a conformal transformation. Cosmologies here
are decelerating, but, only in a particular conformal frame we get eternal acceleration.         
\end{abstract}

\newpage

\noindent{\it 1. Introduction} : S(pace)-like branes are topological defects localized on a space-like hypersurface
which exist as time dependent solutions of many field theories as well as of string/M theory  
\cite{Gutperle:2002ai, Maloney:2003ck}. In string theory just
like D$p$ branes arise as space-like tachyonic kink solution of world volume field theory of non-BPS D$(p+1)$ brane
or D$(p+2)$ -- antiD$(p+2)$ brane \cite{Sen:1999mg}, space-like D$p$ (or SD$p$) branes arise as the 
time-like tachyonic kink solution
of the above unstable brane systems \cite{Sen:2002nu, Gutperle:2002ai}. SD$p$-branes have $(p+1)$ dimensional Euclidean 
world-volume and carry the same
RR charges as their time-like cousins. The original motivation for studying SD$p$-branes was to understand holography
in the temporal context. Just as D$p$ branes give rise to a space-like direction from a Lorentzian world-volume
field theory, SD$p$ branes give rise to a time-like direction from the Euclidean world-volume theory of SD$p$ branes and
this is a necessary ingredient for dS/CFT correspondence \cite{Strominger:2001pn}. One of the reasons for the space-time 
construction of these SD$p$ branes\footnote{The space-time constructions of S-branes were given in \cite{Chen:2002yq,Kruczenski:2002ap,
Roy:2002ik,Bhattacharya:2003sh}.} 
was to understand the so-called dS/CFT correspondence. 

In a previous paper \cite{Roy:2014mba} we constructed an anisotropic (in one direction) SD3 brane solution of type IIB string theory
and compactified on a six dimensional product space of the form $H_5 \times S^1$, where $H_5$ is a five dimensional
hyperbolic space\footnote{Hyperbolic space compactifications are discussed in \cite{Kaloper:2000jb,Starkman:2000dy,Starkman:2001xu}.} 
and $S^1$ is a circle. The resulting 
external space was then shown to be conformal to a four dimensional
de Sitter space. This brought out the connection between SD3 brane and the four dimensional de Sitter space which may
be helpful in understanding dS/CFT correspondence \cite{Strominger:2001pn} in the same spirit as AdS/CFT correspondence
\cite{Maldacena:1997re}. It may be of interest to see
if similar structure exists for other SD$p$ branes for $p \neq 3$. Moreover, it is well-known that S-brane solutions of 
string/M theory give rise to four dimensional accelerating cosmologies (similar to the acceleration of our universe observed in 
the present epoch \cite{Riess:2001gk,Lewis:2002ah,Bennett:2003bz}) upon time dependent hyperbolic space compactification
\cite{Townsend:2003fx,Ohta:2003pu,Roy:2003nd,Emparan:2003gg,Chen:2003dca}
and we have seen this, in particular, for SD2-brane compactified on six dimensional hyperbolic space and expressing the 
resultant metric in Einstein frame \cite{Roy:2003nd,Roy:2006du}. It would be of interest to see whether similar accelerating cosmologies 
can be obtained in other dimensions and under what conditions.      

Motivated by this, we construct in this paper the isotropic SD$p$ brane solutions having isometries ISO$(p+1)$ $\times$ 
SO$(8-p,1)$, from the double Wick rotation of the static, non-supersymmetric, charged D$p$ brane
solutions \cite{Lu:2004ms} of type II string theories. The isotropic SD$p$ brane solutions will be characterized by
three independent parameters ($\tau_0,\,\theta,\,\delta_0$). The parameter $\tau_0$ sets a time scale in the sense that
when $\tau \gg \tau_0$, the solutions become flat. On the other hand when $\tau \sim \tau_0$, the isotropic SD$p$ brane
metrics can be compactified on $(8-p)$ dimensional hyperbolic spaces of time dependent radii, to obtain a $(p+1)+1$ dimensional
flat FLRW metrics in the Einstein frame. We show that these resultant metrics give rise to transient accelerating cosmologies 
for all $p$ (where $1\leq p \leq 6$) i.e., from $(2+1)$ to $(7+1)$ space-time dimensions. The amount of acceleration
and the duration vary with the variations of the various parameters and we study them only in realistic $(3+1)$ space-time dimensions.
When $\tau \ll \tau_0$, we will fix the parameter $\delta_0$ for calculational simplicity (without loss of any generality)
and find after a similar compactification on $(8-p)$ dimensional hyperbolic spaces that the resultant metrics can be
cast into de Sitter forms in $(p+1)+1$ dimensions upto a conformal factor after a suitable coordinate transformation. This 
clarifies the relation
between SD$p$ branes and de Sitter spaces. The other two parameters $\theta$ and $\delta_0$ in the solutions are related to 
the charge of SD$p$ branes and the dilaton, respectively.              

Here we briefly mention that the isotropic SD$p$ brane solutions described in section 2 are not new and they are just rewriting 
the already known solutions \cite{Lu:2004ms,Bhattacharya:2003sh,Lu:2007bu} in a convenient form. The accelerating cosmologies
were known \cite{Ohta:2003pu,Roy:2003nd,Roy:2006du} to follow from the SD2 brane solutions upon time dependent hyperbolic space 
compactification. In this paper we show that similar accelerating cosmologies also follow from all the SD$p$ brane solutions 
(for $1 \leq p \leq 6$) upon similar time dependent hyperbolic space compactification and is described in section 3. The results
of this section are new. Also, a four dimensional de Sitter space upto a conformal factor was obtained before in \cite{Roy:2014mba}
from the near horizon limit of an anisotropic SD3 brane solution of type IIB string theory upon hyperbolic space compactification.
In this paper we show that de Sitter solutions upto a conformal factor in $(p+1)+1$ dimensions also follow from all the isotropic
SD$p$ brane solutions of type II string theory upon hyperbolic space compactification. This is described in section 4 and here
also the results are new. So, the accelerating cosmologies and also the conformally de Sitter solutions are not specific to 
a particular SD$p$ brane (that were known before), but they are quite generic, as we show in this paper, for all the SD$p$ branes 
for $1 \leq p \leq 6$.          

This paper is organized as follows. In the next section, we give the construction of isotropic SD$p$ brane solutions of type
II string theories and write them in a suitable coordinate. In section 3, we show how FLRW type cosmological solutions in
various dimensions can be obtained from the isotropic SD$p$ brane solutions by compatifications. We also discuss about the
solutions in various dimensions. In section 4, we show how the same solutions give rise to $(p+1)+1$ dimensional de Sitter spaces
upto conformal factors in early times. Finally, we conclude in section 5.     

\vspace{.5cm}

\noindent{\it 2. Isotropic SD$p$ brane solutions} :   
In this section we will give the construction of isotropic SD$p$ brane solutions of type II string theories characterized by three
independent parameters and write them in a suitable coordinate system for the ease of our discussion in the next two sections. 
These solutions can actually be obtained either from the static, non-supersymmetric, isotropic $p$-brane solutions in arbitrary space-time
dimensions given in \cite{Lu:2004ms} and using a double Wick rotation, or from the isotropic S-brane solutions in arbitrary dimensions
given in \cite{Bhattacharya:2003sh}. But for convenience we will use the solutions given in eq.(4) of ref.\cite{Lu:2007bu}, representing 
nonsupersymmetric intersecting
brane solutions involving charged D$p$ branes, and chargeless D1 branes and D0 branes. These solutions contain several parameters and to obtain
isotropic nonsupersymmetric D$p$ brane solutions from here we will put the conditions $\delta_2 = \delta_0$, $\bar \delta = (p/4)\delta_0$
and also $\delta_1 = - 2\delta_0$. The solutions eq.(4) of \cite{Lu:2007bu}, then take the form,
\bea\label{dp1}
ds^2 &=& F(r)^{\frac{p+1}{8}} \left(H(r)\tilde{H}(r)\right)^{\frac{2}{7-p}}\left(\frac{H(r)}{\tilde{H}(r)}\right)^{-\frac{p(p+1)}{4(7-p)}\delta_0}\left(dr^2 + 
r^2 d\Omega_{8-p}^2\right)\nn
& & + F(r)^{-\frac{7-p}{8}}\left(\frac{H(r)}{\tilde{H}(r)}\right)^{\frac{p}{4}\delta_0}\left(-dt^2 + \sum_{i=1}^p (dx^i)^2\right)\nn
e^{2(\phi-\phi_0)} &=& F(r)^{\frac{3-p}{2}}\left(\frac{H(r)}{\tilde{H}(r)}\right)^{-(4+p)\delta_0}, \qquad\qquad F_{[8-p]} \,\,=\,\, Q {\rm Vol}(\Omega_{8-p})
\eea 
We remark that the other two references mentioned above also give the same solutions, but the parameter relations are simpler here. Note
that the metrics in \eqref{dp1} are given in the Einstein frame. The various functions appearing in the solutions are defined as,
\bea\label{functions1}
F(r) &=& \left(\frac{H(r)}{\tilde{H}(r)}\right)^{\alpha} \cosh^2 \theta - \left(\frac{\tilde {H}(r)}{H(r)}\right)^{\beta} \sinh^2\theta\nn
H(r) &=& 1 + \frac{\omega^{7-p}}{r^{7-p}},\qquad\qquad \tilde{H}(r)\,\,=\,\, 1 - \frac{\omega^{7-p}}{r^{7-p}}
\eea
There are six parameters $\alpha,\,\beta,\,\delta_0,\,\theta,\,\omega,$ and $Q$ associated with the solutions. However, from the equations
of motion, the parameters can be seen to satisfy the following three relations,
\bea\label{relations1}
& & \alpha - \beta\,\,=\,\, 3\delta_0\nn
& & \frac{14+5p}{7-p}\delta_0^2 + \frac{1}{2}\alpha(\alpha-3\delta_0)\,\,=\,\, \frac{8-p}{7-p}\nn
& & Q \,\,=\,\, (7-p) \omega^{7-p}(\alpha+\beta)\sinh2\theta
\eea
Using these relations we can eliminate three parameters out of the six we mentioned above and therefore, the solutions have three independent
parameters, namely, $\omega,\,\theta$ and $\delta_0$. Note from the form of $\tilde{H}(r)$ in \eqref{functions1} that the solutions have  
curvature singularities at $r=\omega$ and therefore, the solutions are well defined only for $r>\omega$. Also in \eqref{dp1} $\phi_0$ denotes the
asymptotic value of the dilaton and $F_{[8-p]}$ is the $(8-p)$ form and $Q$ is the charge associated with the D$p$ branes which in this case
are magnetically charged. We point out that the singularities at $r=\omega$ are naked singularities where the dilaton becomes plus or minus
infinity (depending on the parameters and the value of $p$) and can not be removed by coordinate transformations
or going to a different (dual) frame. However, since these are string theory solutions it may be plausible that these singularities will go away when
the higher order stringy effects are taken into account. As far as we know the status of these singularities in full string theory is still not clearly
understood.   

Now in order to get isotropic SD$p$ brane solutions we apply the double Wick rotation \cite{Lu:2004ms} $r \to it$, $t \to -ix^{p+1}$ to the 
solutions \eqref{dp1}
along with $\omega \to i\omega$, $\theta \to i\theta$ and $\theta_1 \to i\theta_1$, where $\theta_1$ is one of the angles parameterizing the sphere
$\Omega_{8-p}$ and then we obtain,
\bea\label{sdp1}
ds^2 &=& F(t)^{\frac{p+1}{8}} \left(H(t)\tilde{H}(t)\right)^{\frac{2}{7-p}}\left(\frac{H(r)}{\tilde{H}(r)}\right)^{-\frac{p(p+1)}{4(7-p)}\delta_0}\left(-dt^2 + 
t^2 dH_{8-p}^2\right)\nn
& & + F(t)^{-\frac{7-p}{8}}\left(\frac{H(t)}{\tilde{H}(t)}\right)^{\frac{p}{4}\delta_0}\sum_{i=1}^{p+1} (dx^i)^2\nn
e^{2(\phi-\phi_0)} &=& F(t)^{\frac{3-p}{2}}\left(\frac{H(t)}{\tilde{H}(t)}\right)^{-(4+p)\delta_0}, \qquad\qquad F_{[8-p]} \,\,=\,\, (-1)^{8-p} Q {\rm Vol}(H_{8-p})
\eea 
where the various functions are now given as,
\bea\label{functions2}
F(t) &=& \left(\frac{H(t)}{\tilde{H}(t)}\right)^{\alpha} \cos^2 \theta + \left(\frac{\tilde {H}(t)}{H(t)}\right)^{\beta} \sin^2\theta\nn
H(t) &=& 1 + \frac{\omega^{7-p}}{t^{7-p}},\qquad\qquad \tilde{H}(t)\,\,=\,\, 1 - \frac{\omega^{7-p}}{t^{7-p}}
\eea
Note that under the Wick rotation the solutions have become time dependent. The naked singularities at $r=\omega$ has now changed to the
singularities at $t=\omega$. Also, the metric of the sphere $d\Omega_{8-p}^2$ has changed to negative
of the metric of the hyperbolic space $dH_{8-p}^2$. The metrics now has the symmetry ISO$(p+1)$ $\times$ SO$(8-p,1)$. The hyperbolic functions
$\sinh\theta$, $\cosh\theta$ have become trigonometric functions and the function $F$ has relative plus sign in the two terms instead of minus.
Most importantly the form field remains real and retains its form upto a sign which does not happen for the BPS D$p$ branes (Wick rotation 
actually makes the form field imaginary for BPS D$p$ branes and the solutions in that case do not remain solutions of type II theories, instead
they become solutions of type II$^{\ast}$ theories \cite{Hull:2001ii}). The first two parameter relations in \eqref{relations1} 
remain the same under the
Wick rotation, whereas the last relation changes to $Q = (7-p) \omega^{7-p}(\alpha+\beta)\sin2\theta$ if we insist that $Q$ should also change
under the Wick rotation as $Q \to (i)^{8-p} Q$. Eq.\eqref{sdp1} represents real isotropic SD$p$ brane solutions of type II string theories characterized by
three independent parameters $\omega,\,\theta,\,\delta_0$. 

Now for the discussion in the next two sections we will make a coordinate transformation from $t$ to $\tau$ given by,
\be\label{trans}
t \,\,=\,\, \tau\left(\frac{1+\sqrt{g(\tau)}}{2}\right)^{\frac{2}{7-p}}, \qquad {\rm where,}\qquad g(\tau)\,\, =\,\, 1+\frac{4\omega^{7-p}}{\tau^{7-p}} \equiv
1 + \frac{\tau_0^{7-p}}{\tau^{7-p}}
\ee
Under this coordinate change we get,
\bea\label{fnschange}
& & H(t) = 1 + \frac{\omega^{7-p}}{t^{7-p}} = \frac{2\sqrt{g(\tau)}}{1+\sqrt{g(\tau)}}, \qquad \tilde{H}(t) =  1 - \frac{\omega^{7-p}}{t^{7-p}} 
= \frac{2}{1+\sqrt{g(\tau)}},\nn
& & H(t)\tilde{H}(t) = \frac{4\sqrt{g(\tau)}}{(1+\sqrt{g(\tau)})^2}, \qquad \frac{H(t)}{\tilde{H}(t)} = \sqrt{g(\tau)},\nn
& & -dt^2 + t^2 dH_{8-p}^2 = g(\tau)^{\frac{1}{7-p}}\left(-\frac{d\tau^2}{g(\tau)} + \tau^2 dH_{8-p}^2\right)
\eea 
Using these relations we can rewrite the isotropic SD$p$ brane solutions given in \eqref{sdp1} as follows,
\bea\label{sdp2}
ds^2 &=& F(\tau)^{\frac{p+1}{8}} g(\tau)^{\frac{1}{7-p} - \frac{p(p+1)}{8(7-p)}\delta_0}\left(-\frac{d\tau^2}{g(\tau)} + 
\tau^2 dH_{8-p}^2\right)
+ F(\tau)^{-\frac{7-p}{8}} g(\tau)^{\frac{p}{8}\delta_0}\sum_{i=1}^{p+1} (dx^i)^2\nn
e^{2(\phi-\phi_0)} &=& F(\tau)^{\frac{3-p}{2}} g(\tau)^{-\frac{(4+p)}{2}\delta_0}, \qquad\qquad F_{[8-p]} \,\,=\,\, (-1)^{8-p} Q {\rm Vol}(H_{8-p})
\eea   
where $g(\tau)$ is as given in \eqref{trans} and $F(\tau)$ is given by,
\be\label{ftau}
F(\tau) = g(\tau)^{\frac{\alpha}{2}} \cos^2\theta + g(\tau)^{-\frac{\beta}{2}} \sin^2\theta
\ee
The parameter relations remain the same as given in \eqref{relations1} with the factor $\sinh2\theta$ in the last one replaced 
by $\sin2\theta$. It should be noted from \eqref{sdp2}, that in the new coordinate, the original singularities at $t=\omega$ have been 
shifted to $\tau=0$. Now the solutions have three independent parameters, namely, $\tau_0,\,\theta,\,\delta_0$. Also note that as $\tau \gg \tau_0$,
$g(\tau),\, F(\tau) \to 1$ and therefore, the solutions reduce to flat space. In the next two sections we will use the solutions \eqref{sdp2}
to see how one can get cosmologies in various dimensions and also how to obtain de Sitter spaces upto a conformal factor.

\vspace{.5cm}

\noindent{\it 3. FLRW cosmologies from SD$p$ brane compactifications} : In this section we will see how we can get flat FLRW cosmologies
in various dimensions from the isotropic SD$p$ brane solutions given in \eqref{sdp2}.
We will assume that $\tau \sim \tau_0$, so that the two terms in the function $g(\tau) = 1 + \tau_0^{7-p}/\tau^{7-p}$ are comparable
and we must keep both the terms. Keeping this in mind we can rewrite the metrics in \eqref{sdp2} in the following form,
\be\label{flrw1}
ds^2 = F(\tau)^{-\frac{(p+1)(8-p)}{8p}} g(\tau)^{-\frac{8-p}{p(7-p)}+\frac{(8-p)(p+1)}{8(7-p)}\delta_0} \tau^{-\frac{2(8-p)}{p}} ds_E^2 
+ F(\tau)^{\frac{p+1}{8}} g(\tau)^{\frac{1}{7-p} - \frac{p(p+1)}{8(7-p)}\delta_0} \tau^2 dH_{8-p}^2
\ee
where
\be\label{flrw2}
ds_E^2 = -  F(\tau)^{\frac{p+1}{p}} g(\tau)^{\frac{8}{p(7-p)}-\frac{p+1}{7-p}\delta_0 - 1} \tau^{\frac{2(8-p)}{p}}\,d\tau^2
+  F(\tau)^{\frac{1}{p}} g(\tau)^{\frac{8-p}{p(7-p)}-\frac{1}{7-p}\delta_0} \tau^{\frac{2(8-p)}{p}} \sum_{i=1}^{p+1} (dx^i)^2
\ee  
is the $(p+1)+1$ dimensional metrics in the Einstein frame. One can think of these metrics as coming from the compactification of the ten
dimensional metrics \eqref{flrw1} on $(8-p)$ dimensional hyperbolic space with time dependent radius given by 
\be\label{radius}
R(\tau) = F(\tau)^{\frac{p+1}{16}} g(\tau)^{\frac{1}{2(7-p)} - \frac{p(p+1)}{16(7-p)}\delta_0} \tau
\ee 
and expressing the resulting metrics in the Einstein frame. Now defining a new time coordinate $\eta$ by,
\be\label{newtime}
d\eta = F(\tau)^{\frac{p+1}{2p}} g(\tau)^{\frac{4}{p(7-p)}-\frac{p+1}{2(7-p)}\delta_0 - \frac{1}{2}} \tau^{\frac{8-p}{p}}\,d\tau
\equiv C(\tau) d\tau
\ee
we can rewrite the Einstein frame metrics $ds_E^2$ in the standard flat FLRW form in $(p+1)+1$ dimensions as
\be\label{flrw3}
ds_E^2 = - d\eta^2 + S^2(\eta)\sum_{i=1}^{p+1} (dx^i)^2
\ee
where the scale factor $S(\eta)$ is given by,
\be\label{sf}
S(\eta) \equiv A(\tau) =  F(\tau)^{\frac{1}{2p}} g(\tau)^{\frac{8-p}{2p(7-p)}-\frac{1}{2(7-p)}\delta_0} \tau^{\frac{8-p}{p}}
\ee
Now we define another function
\be\label{btau}
B(\tau) = \frac{A(\tau)}{C(\tau)} = F(\tau)^{-\frac{1}{2}} g(\tau)^{\frac{6-p}{2(7-p)} + \frac{p}{2(7-p)}\delta_0}
\ee
where $C(\tau)$ is defined in \eqref{newtime}. Now the universe is expanding if the scale factor $S(\eta)$ satisfies
$dS(\eta)/d\eta > 0$ and the expansion is accelerating if it further satisfies $d^2 S(\eta)/d\eta^2 > 0$. Since $S(\eta)$
is a complicated function of $\eta$, we will translate \cite{Roy:2006du} these two conditions in terms of the two known 
functions $A(\tau)$ and $B(\tau)$ given in \eqref{sf} and \eqref{btau}. The conditions are,
\bea\label{cond}
m(\tau) &\equiv & \frac{d\ln A(\tau)}{d\tau} > 0\nn
n(\tau) &\equiv & \frac{d^2\ln A(\tau)}{d\tau^2} + \frac{d\ln A(\tau)}{d\tau} \frac{d\ln B(\tau)}{d\tau} > 0
\eea
where $m(\tau)$ is the expansion parameter and $n(\tau)$ is the rate of expansion parameter. The parameters $\alpha$ and
$\beta$ which appear in the definition of $F(\tau)$ given in \eqref{ftau} can be given in terms of $\delta_0$ from the
second relation in \eqref{relations1} as,
\bea\label{alphabeta}
\alpha &=& \frac{3}{2}\delta_0 \pm \sqrt{\frac{8(8-p) - 49(p+1)\delta_0^2}{4(7-p)}}\nn
\beta &=& -\frac{3}{2}\delta_0 \pm \sqrt{\frac{8(8-p) - 49(p+1)\delta_0^2}{4(7-p)}}
\eea

\begin{figure}[ht]
\begin{center}
\includegraphics[width=0.5\textwidth,height=6cm]{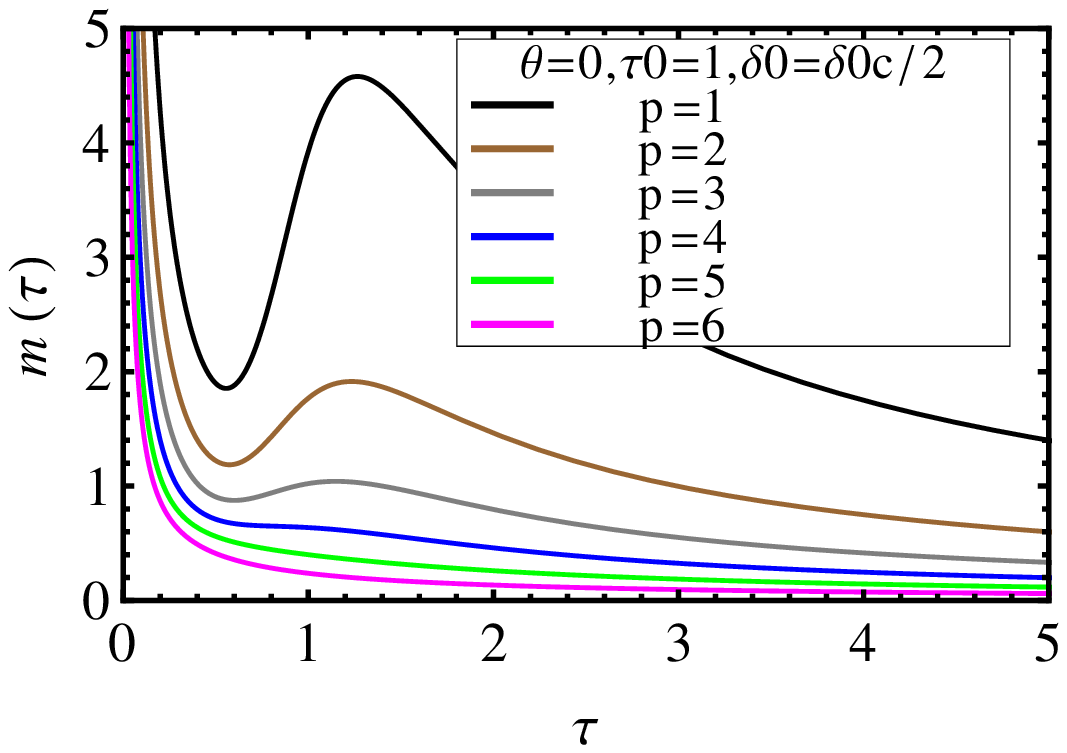}\includegraphics[width=0.5\textwidth,height=6cm]{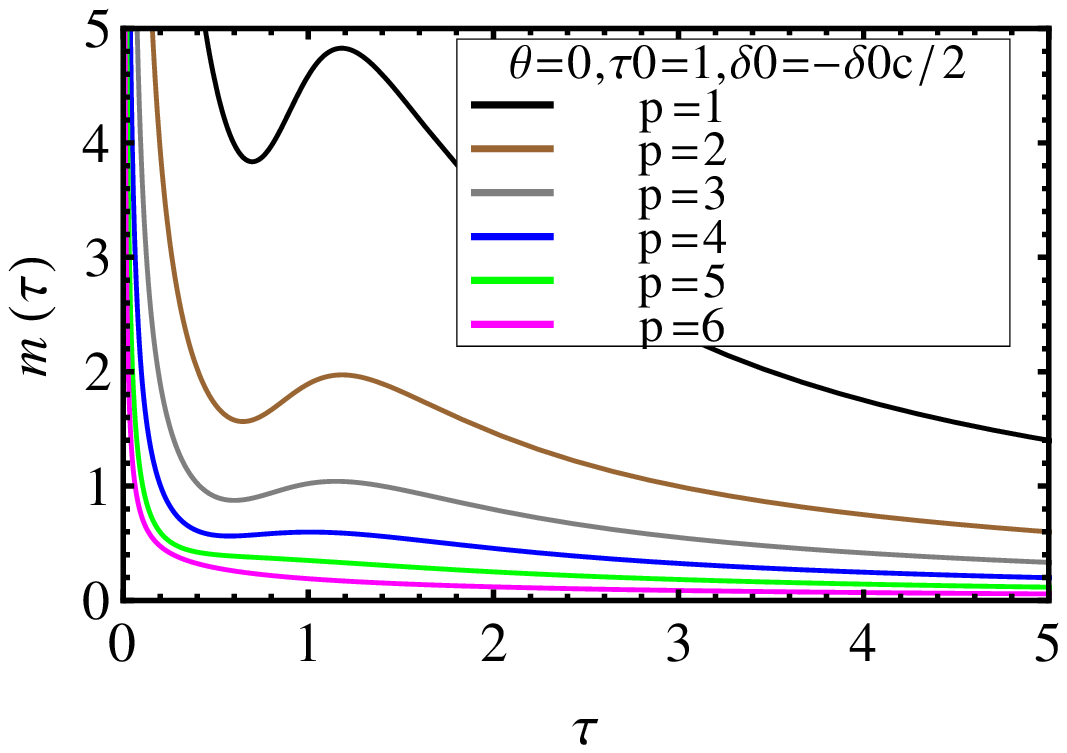}
\end{center}
\caption[\textwidth]{Plot of expansion paramater $m(\tau)$ for SD$p$ brane compactifications on hyperbolic space 
$H_{8-p}$ given in 
\eqref{cond}. The functions are plotted for $\theta=0$, $\tau_0=1$, $\delta_0=\frac{\delta_{0c}}{2} = \frac{1}{14}\sqrt{\frac{2(8-p)}{p+1}}$ and 
$p=1,\ldots,6$ in the left panel and for $\theta=0$, $\tau_0=1$, $\delta_0=-\frac{\delta_{0c}}{2} = -\frac{1}{14}\sqrt{\frac{2(8-p)}{p+1}}$ and 
$p=1,\ldots,6$ in the right panel.}
\end{figure}

\begin{figure}[ht]
\begin{center}
\includegraphics[width=0.5\textwidth,height=6cm]{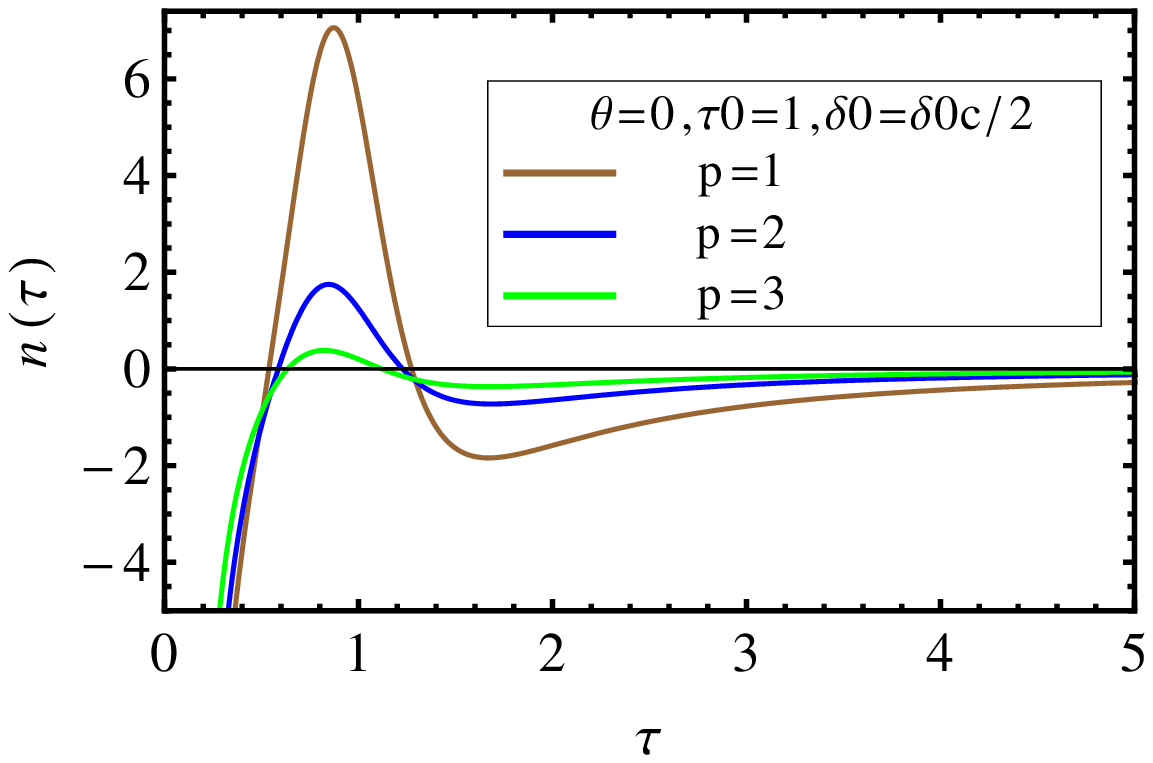}\includegraphics[width=0.5\textwidth,height=6cm]{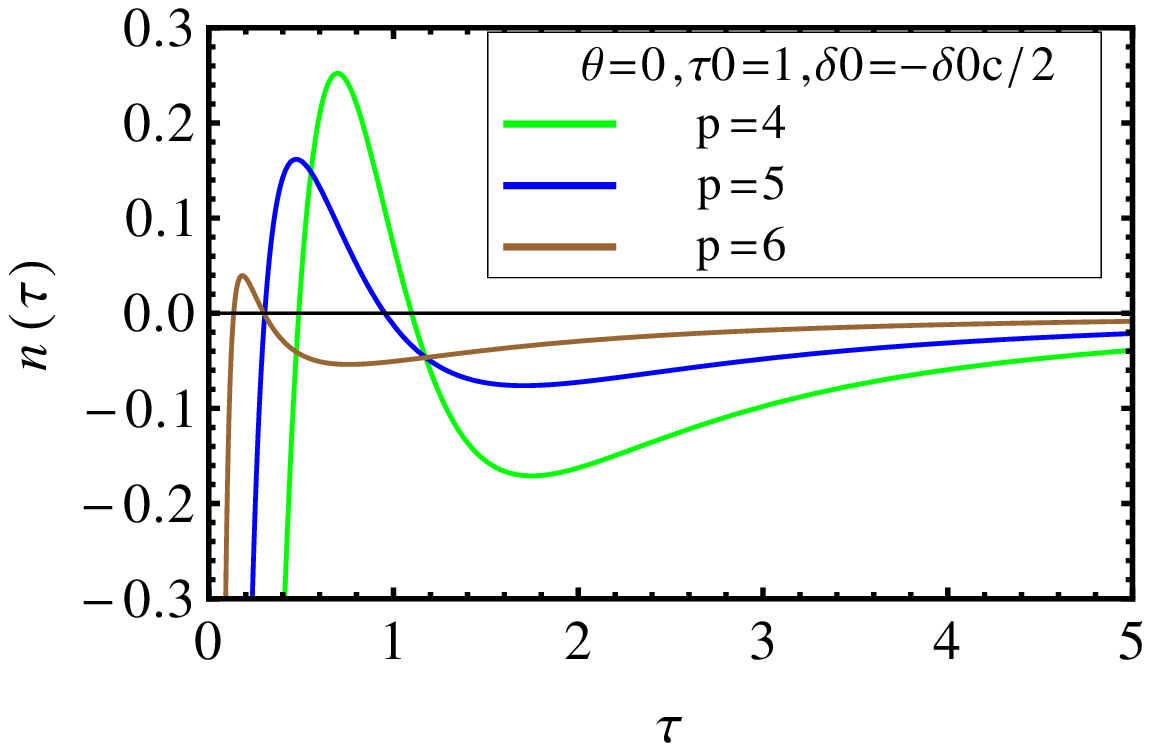}
\end{center}
\caption[\textwidth]{Plot of rate of expansion parameter $n(\tau)$ for SD$p$ brane compactifications on hyperbolic space $H_{8-p}$ given in
\eqref{cond}. Here the functions are plotted for $\theta=0$, $\tau_0=1$, $\delta_0=\frac{\delta_{0c}}{2}=
\frac{1}{7}\sqrt{\frac{2(8-p)}{p+1}}$ and 
$p=1,2,3$ in the left panel and $\theta=0$, $\tau_0=1$, $\delta_0=-\frac{\delta_{0c}}{2}=
-\frac{1}{7}\sqrt{\frac{2(8-p)}{p+1}}$ and $p=4,5,6$ in the right panel.}
\end{figure}

In Figures 1, 2, we have plotted the expansion parameter $m(\tau)$ and the rate of expansion parameter $n(\tau)$, 
respectively. 
We have used the $(p+1)+1$ dimensional metrics $ds_E^2$ given in \eqref{flrw2} and the functions $A(\tau)$ and $B(\tau)$ given in 
\eqref{sf} and \eqref{btau}. We have also used the value of $\alpha$, $\beta$ given in \eqref{alphabeta}. 
The positive value of $m(\tau)$ in Figure 1 indicates the expansion of the universe. From the 
above plot, we see that the universe expands for all values of $p$ (where $1 \leq p \leq 6$) and therefore, we get the expanding $2+1$, $3+1$ upto 
$7+1$ dimensional universes. In Figure 1, the values of the various parameters we have chosen are $\theta=0$ (this means that the form field
is zero and therefore the solution is chargeless and simpler), $\tau_0=1$ (this is a typical value we have chosen to show the cosmologies
in various dimensions and if $\tau_0$ is less than this value the acceleration is more but the duration is less as seen in Figure 5) 
and $\delta_0=\delta_{0c}/2$ (defined below) in the left panel and $\delta_0 = -\delta_{0c}/2$ in the right panel. Actually, the parameter $\delta_0$ can not
take any arbitrary value. From \eqref{alphabeta} we note that since the parameters $\alpha$ and $\beta$ are real $\delta_0$ must lie in between
\be\label{delta0}
-\frac{2}{7}\sqrt{\frac{2(8-p)}{p+1}} \leq \delta_0 \leq \frac{2}{7}\sqrt{\frac{2(8-p)}{p+1}} = \delta_{0c},
\ee
where we have called the maximum value of $\delta_0$ as $\delta_{0c}$. In Figure 1, we have chosen the value of $\delta_0$ as $\pm 1/2$
of its maximum value in the left panel and in the right panel respectively and get expanding universes in all dimensions. The reason for choosing
these particular values is that we get accelerating expansion for these values for different $p$ as shown in Figure 2.  
Figure 2 also contains two panels. Here again the positivity of $n(\tau)$ gives an accelerating phase of expansion. 
On the left panel of Figure 2, we show that $n(\tau)$ remains positive for certain interval of time for $p=1,\,2,\,3$ and on the right panel
we show the positivity of $n(\tau)$ for certain interval of time for $p=4,\,5,\,6$. Therefore, we get accelerating expansions for all
values of $p$ from 1 to 6. Note that the magnitude of acceleration and the duration depend crucially on the parameters $\theta$, $\tau_0$ and
particularly $\delta_0$. If the parameters are not chosen judiciously, we do not get accelerations. For $p=1,\,2,\,3$, we have chosen $\theta=0$,
$\tau_0=1$ and $\delta_0 = \delta_{0c}/2$ in the left panel of Figure 2 to get accelerations. If we keep the same values of the parameters we get 
acceleration for
$p=4$ but no accelerations for $p=5,\,6$. This is the reason, for $p=4,\,5,\,6$ we have chosen $\theta=0$, $\tau_0=1$ and $\delta_0 = -\delta_{0c}/2$
in the right panel of Figure 2 and get accelerations in all the cases. This shows that we can get accelerating cosmologies for all values of $p$
by the appropriate choice of the various parameters characterizing the SD$p$ solutions.      
The expansion, however, becomes decelerating in the remote past, i.e., for $\tau \ll \tau_0$ and also
in the far future $\tau \gg \tau_0$ irrespective of spacetime dimensions and other parameters and all the curves tend to merge in those two regions.
We will discuss those cases later.
\begin{table}
\centering
\begin{tabular}{|c|c|c|c|}
\hline
\hline
$(p+1)+1$ & $\delta_0$ & $\tau$ & $n(\delta_0,\tau)$\\
\hline
2+1 & $\frac{\delta_{0c}}{2}$ & 0.87197 & 7.05882\\
\hline 
3+1 & $\frac{\delta_{0c}}{4}$ & 0.84025 & 1.93012\\
\hline 
4+1 & $ 0 $ & 0.78463 & 0.72178\\
\hline 
5+1 & $-\frac{\delta_{0c}}{4}$ & 0.68025 & 0.30926\\
\hline 
6+1 & $-\frac{\delta_{0c}}{2}$ & 0.47336 & 0.16178\\
\hline 
7+1 & $-\frac{3\delta_{0c}}{4}$ & 0.11068 & 0.32242\\
\hline 
\hline
\end{tabular}
\caption{Here $n(\delta_0,\tau)$ is treated as a function of two variables $\delta_0$ and $\tau$. We have found the 
particular value of $\delta_0$ which makes the rate of expansion maximum for $\theta=0$, $\tau_0=1$ and $p=1,\ldots,6$.}
\end{table}
We have tabulated the values of $\delta_0$ for which the rate of expansion parameter $n(\tau)$ is maximum for different values
of $p$ in Table 1. We have also given those maximum values and the values of $\tau$ where these maxima occurs. We have chosen $\theta=0$ and $\tau_0=1$. 
In all the cases the maximum values are found to be positive and so there are accelerations for all $p$. 

\begin{figure}[ht]
\begin{center}
\includegraphics[width=0.75\textwidth,height=8cm]{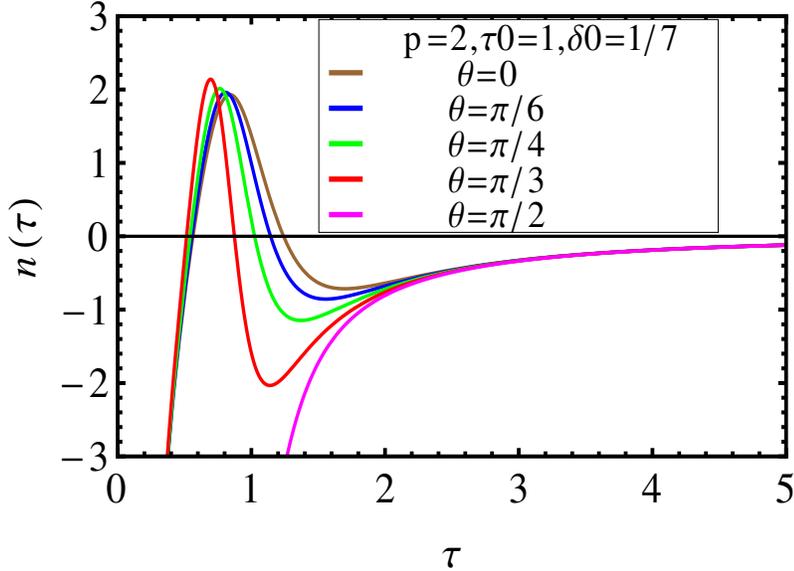}
\end{center}
\caption[\textwidth]{The plot of $n(\tau)$ for different values of $\theta$ where $p=2$, $\tau_0=1$ and $\delta_0=\frac{\delta_{0c}}{4}= 
\frac{1}{7}$.}
\end{figure}

\begin{figure}
\begin{center}
\includegraphics[width=0.75\textwidth,height=8cm]{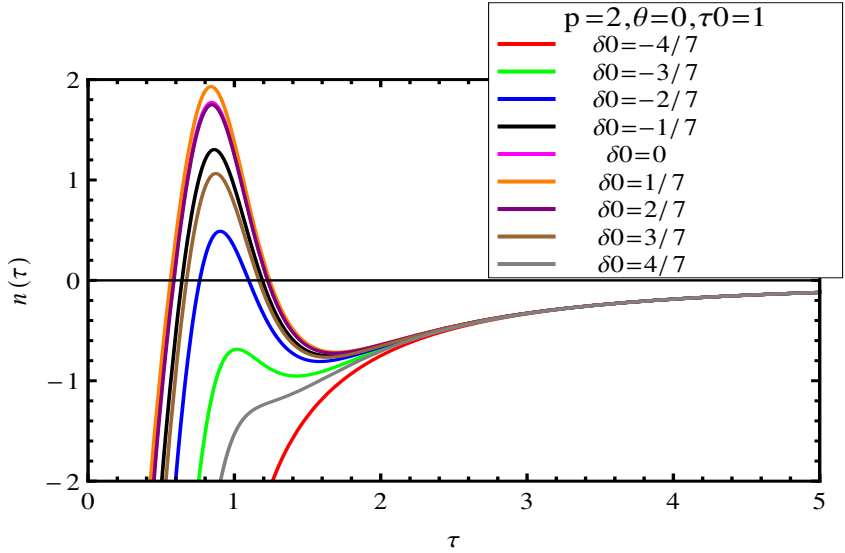}
\end{center}
\caption[\textwidth]{The plot of $n(\tau)$ for $\delta_0= -\delta_{0c} (= -\frac{4}{7})$, $-\frac{3\delta_{0c}}{4} 
(= -\frac{3}{7})$, $-\frac{\delta_{0c}}{2} (= -\frac{2}{7})$, $-\frac{\delta_{0c}}{4} (= -\frac{1}{7})$, 0, $\frac{\delta_{0c}}{4} 
(= \frac{1}{7})$, $\frac{\delta_{0c}}{2} (= \frac{2}{7})$, $\frac{3\delta_{0c}}{4} (= \frac{3}{7})$, $\delta_{0c} (= \frac{4}{7})$, $p=2$ 
and $\tau_0=1$.}
\end{figure}
\begin{figure}[ht]
\begin{center}
\includegraphics[width=0.75\textwidth,height=8cm]{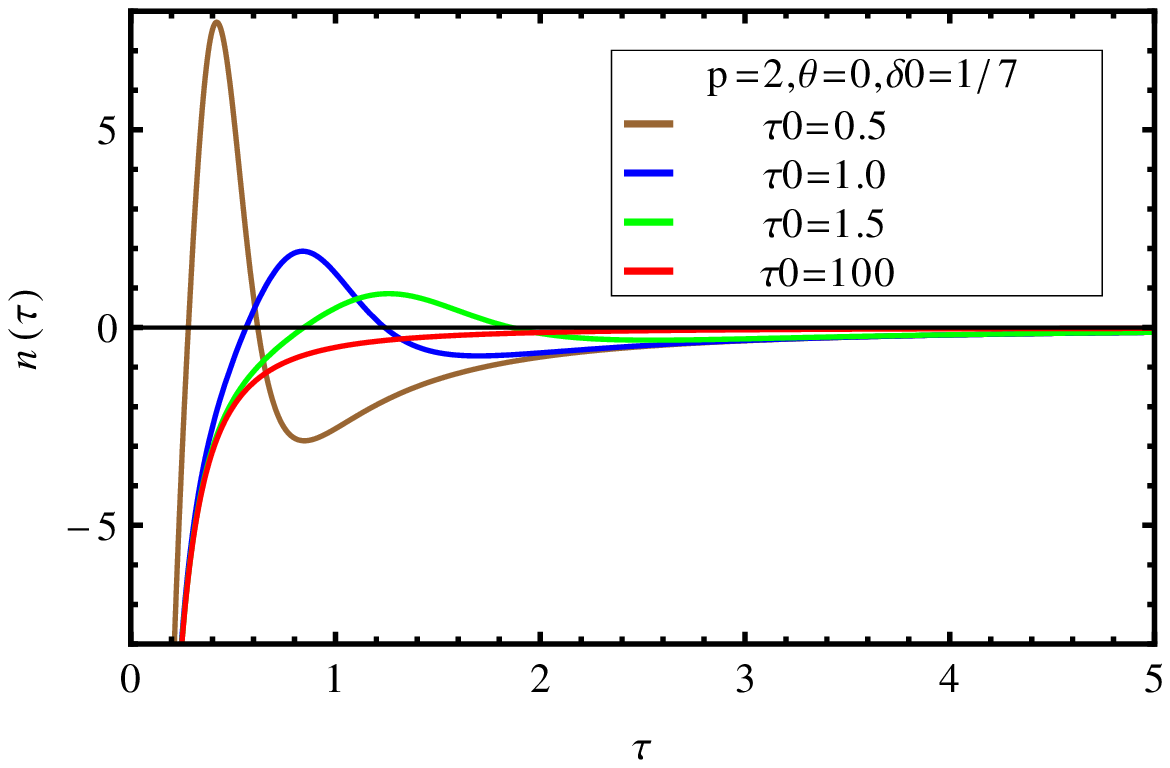}
\end{center}
\caption[\textwidth]{Plot of $n(\tau)$ for different values of $\tau_0$, with the other parameters kept fixed at $\theta=0$, 
$\delta_0=\frac{\delta_{0c}}{4}=\frac{1}{7}$ and $p=2$}
\end{figure}

In Figures 3, 4, 5 below, we have plotted the rate of expansion parameter $n(\tau)$ for various values of $\theta$, the charge parameter, $\delta_0$,
the dilaton parameter and $\tau_0$, the time scale, respectively. We have
taken $p=2$, so that the space-time is $3+1$ dimensional. In Figure 3, we have taken $\tau_0=1$ and $\delta_0=\delta_{0c}/4=1/7$.
We find that there is acceleration for all values of $\theta$ in the range $0 \leq \theta < \pi/2$. The acceleration is 
minimum for $\theta=0$ and it gradually increases as we increase the value of $\theta$, except at $\theta=\pi/2$. 
The duration of the accelerating phase gradually decreases with increasing $\theta$ and becomes zero exactly at $\frac{\pi}{2}$.
This happens for every dimension where there is an accelerating phase. Note that here we have used the upper sign of $\alpha$ given in
\eqref{alphabeta}. If we use the lower sign we get exactly the same behavior with the interchange of $\theta=0$ and $\theta=\pi/2$.
Similar results can also be obtained for other values of $p$. 
In Figure 4, we have plotted the rate of expansion parameter $n(\tau)$ for different values of $\delta_0$, with the other parameters kept  
fixed at $\theta=0$ and $\tau_0=1$. We have again chosen $p=2$, corresponding to $3+1$ dimensional universe. The accelerating phase 
depends on the value of $\delta_0$. We varied $\delta_0$ from $\delta_{0c}$ to $-\delta_{0c}$ in an interval of $(1/4)\delta_{0c}$. We find 
from the figure that the acceleration is maximum for $\delta_0=\frac{\delta_{0c}}{4}$. Acceleration decreases
for other absolute values of $\delta_0$. For $\delta_0 = 0$ and $ \delta_{0c}/2$ we get reasonably large acceleration although plus
value gives more acceleration than minus value. On the other hand, for $\delta_0 = \pm \delta_{0c}$ we always get deceleration. Again
this happens for each dimension where there is an accelerating phase. As before there exists a critical absolute value of $\delta_0$ for
each $p$ above which there is an accelerating phase and below which there will always be deceleration. Similar conclusion can be drawn for
other values of $p$. 
In Figure 5, we have plotted $n(\tau)$ for different values of $\tau_0$, with the other parameters kept fixed at $\theta=0$ and 
$\delta_0=\delta_{0c}/4$. We have chosen $p=2$ as before such that we have $3+1$ dimensional space-time. Here we find that the acceleration
is more but of shorter duration as we decrease the value of $\tau_0$. However, as $\tau_0$ is increased beyond a certain value we always have 
deceleration. This also happens for each dimension where there is an accelerating phase. There is a critical value of $\tau_0$ at each
dimension $p=1,\ldots,6$, below which there is an accelerating phase and above which there is always a deceleration.

We remark that even though it is difficult to obtain an exact relation between $\eta$ and $\tau$ from the relation \eqref{newtime}, but
it can be integrated in the far future $\tau \gg \tau_0$ and also in the remote past $\tau \ll \tau_0$. The case of $\tau \ll \tau_0$ will
be discussed in the next section. Here we mention that for $\tau \gg \tau_0$, $\tau$ is related to $\eta$ by the relation $\tau \sim 
(\eta-\eta_0)^{\frac{p}{8}}$. Therefore, we have the scale factor \eqref{sf} to take the form $S(\eta) \sim (\eta-\eta_0)^{1-\frac{p}{8}}$. It is
clear from here that in the far future the universe will expand with deceleration for all $p$ as we have seen in Figures 1,2.
 
\vspace{.5cm}

\noindent{\it 4. Conformally de Sitter spaces in various dimensions} : 
In this section we will see how at early times $\tau \ll \tau_0$, we can get de Sitter solutions upto a conformal transformation
in various dimensions from the SD$p$ solutions given in \eqref{sdp2}. Note that in this case the function $g(\tau)$ can be 
approximated as,
\be\label{gtau} 
g(\tau)= 1 + \frac{\tau_0^{7-p}}{\tau^{7-p}} \approx \frac{\tau_0^{7-p}}{\tau^{7-p}}
\ee
and so the function $F(\tau)$ given in \eqref{ftau} can be approximated as,
\be\label{ftau1}
F(\tau) \approx \left(\frac{\tau_0}{\tau}\right)^{\frac{7-p}{2}\alpha} \cos^2\theta
\ee   
Here we have assumed $\theta \neq \pi/2$, otherwise, it is arbitrary. If $\theta=\pi/2$, then $F(\tau)$ takes the form
$F(\tau) = (\frac{\tau_0}{\tau})^{-(7-p)\beta/2}$. But, as we will see that since the final answer will be independent of the parameters
$\alpha,\,\beta$, we can take the form of $F(\tau)$ as given in \eqref{ftau1} without any loss of generality. We will further choose
$\alpha+\beta=2$ for calculational simplicity and again without losing any generality. Now since from the parameter relations given
in \eqref{relations1} we have $\alpha-\beta=3\delta_0$, combining these two we get $\alpha=1+(3/2)\delta_0$. For more simplification we
will set $\tau_0=1$ and $\theta=0$. With all these the metric and the dilaton in \eqref{sdp2} take the forms,
\bea\label{dsmetric1}
ds^2 &=& -\,\tau^{\frac{(7-p)(15-p)-16}{16} - \frac{7(3-p)(p+1)}{32}\delta_0} d\tau^2 + \tau^{\frac{(7-p)^2}{16} + \frac{7(7-p)(3-p)}{32}\delta_0} \sum_{i=1}^{p+1}
(dx^i)^2\nn 
& & \qquad\qquad\qquad\qquad\qquad +\, \tau^{\frac{16-(7-p)(p+1)}{16} - \frac{7(3-p)(p+1)}{32}\delta_0} dH_{8-p}^2\nn
e^{2(\phi-\phi_0)} &=& \tau^{-\frac{(3-p)(7-p)}{4} + \frac{7(7-p)(p+1)}{8}\delta_0}
\eea  
It should be mentioned that here $\delta_0$ is not a free parameter, unlike in the previous section where we did not use
$\alpha+\beta=2$. In fact, since $\alpha = 1+(3/2)\delta_0$, we can use the second
parameter relation in \eqref{relations1} to obtain the value of $\delta_0$ as,
\be\label{delta_0}
\delta_0 = \pm \frac{2}{7}\sqrt{\frac{9-p}{p+1}}
\ee
Now the metrics in \eqref{dsmetric1} can also be written as,
\be\label{dsmetric2}
ds^2 = -\,\left(\tau^{\frac{16-(7-p)(p+1)}{16} - \frac{7(3-p)(p+1)}{32}\delta_0}\right)^{\frac{p-8}{p}} ds_E^2
+\, \tau^{\frac{16-(7-p)(p+1)}{16} - \frac{7(3-p)(p+1)}{32}\delta_0} dH_{8-p}^2
\ee
where $ds_E^2$ is a $(p+1)+1$ dimensional metrics in the Einstein frame and have the forms,
\be\label{dsmetric3}
ds_E^2 = \tau^{\frac{(9-p)(p+1)}{2p} - \frac{7(3-p)(p+1)}{4p}\delta_0} \left[-\frac{d\tau^2}{\tau^2} + \tau^{-\frac{(9-p)}{2} + \frac{7(3-p)}{4}\delta_0} \sum_{i=1}^{p+1}
(dx^i)^2\right]
\ee
Actually the metrics in \eqref{dsmetric3} can be seen to arise from a $(8-p)$ dimensional hyperbolic space compactifications
with time dependent radius $R(\tau)=\tau^{\frac{16-(7-p)(p+1)}{8} - \frac{7(3-p)(p+1)}{16}\delta_0}$ and then expressing the resulting $(p+1)+1$ 
dimensional metrics in the Einstein frame. We notice that for $p=3$, $R(\tau)=1$ and for this case the transverse hyperbolic space $H_5$
gets decoupled from the rest of the space-time, similar to what happens for D3 brane where the transverse $S^5$ gets decoupled. This 
simplification for $p=3$ case occurs because of our particular choice of parameters, namely, $\alpha+\beta=2$.  
Defining a canonical time by the relation,
\be\label{cantime}
\eta^2 = \tau^{\frac{(9-p)}{2} - \frac{7(3-p)}{4}\delta_0}
\ee
we can rewrite the metrics in \eqref{dsmetric3} as,
\be\label{dsmetric4}
ds_E^2 = \eta^{\frac{2(p+1)}{p}}\left[-\frac{d\eta^2}{\eta^2} + \frac{\sum_{i=1}^{p+1}(dx^i)^2}{\eta^2}\right]
\ee  
Note that in writing the above metrics we have scaled $\eta$ and $x^i$'s as follows,
\bea\label{scaling}
\eta & \to & \left(\frac{9-p}{4} - \frac{7(3-p)}{16} \delta_0\right)^{\frac{p}{p+1}} \eta\nn
x^i & \to & \left(\frac{9-p}{4} - \frac{7(3-p)}{16}\delta_0\right)^{\frac{p}{p+1}} x^i, \qquad {\rm for} \quad i=1,\ldots,(p+1)
\eea
The dilaton given in \eqref{dsmetric1} can also be written in terms of canonical time using \eqref{cantime}. We recognize
the metrics in \eqref{dsmetric4} to be the de Sitter metrics in $(p+1)+1$ dimensions upto the conformal factor
$\eta^{2(p+1)/p}$. For $p=2$, i.e., for the four dimensional case the conformal factor becomes $\eta^3$ precisely the form
we obtained in \cite{Roy:2014mba}. 

We can compare the situation here with the time-like or static BPS D$p$-brane cases. For the usual D$p$ branes when $p=3$,
the near horizon limit gives AdS$_5$ $\times$ S$^5$ solution. In other words, compactifying on S$^5$, we get AdS$_5$ solution.
So, here the boundary theory is conformally invariant. But for other D$p$ branes (except for $p=5$), compactifying on S$^{8-p}$,
we do not get AdS$_{p+2}$ spaces directly, but we get them upto a conformal factor. So, the boundary theories for these cases do 
not have conformal symmetry (as the bulk spaces are not AdS spaces -- conformal factors make the bulk spaces to be different
from AdS spaces), but still the connection with the AdS spaces upto a conformal transformation proves to be useful for calculational 
purposes. For, space-like D$p$ branes compactifying on hyperbolic spaces H$_{8-p}$, we never get (for any $p$) de Sitter spaces, but we get
them (dS$_{p+2}$) upto a conformal factor exactly like static D$p$ branes with $p \neq 3,\,5$. Here also the boundary theories
do not have conformal symmetry since the bulk solution is not a de Sitter solution. However, this connection with de Sitter solution
with the space-like D$p$ branes might prove to be useful for calculational purposes (for example, calculation of correlation function)
as in the static D$p$ brane cases.   
 
To see that the space-times given in \eqref{dsmetric4} describe decelerating expansions we rewrite them in flat FLRW forms
by defining a new canonical coordinate by $d\tilde\eta = \eta^{\frac{1}{p}}d\eta$. The metrics in \eqref{dsmetric4} then takes
the forms,
\be\label{dsmetric5}
ds_E^2 = -d\tilde\eta^2 + S^2(\tilde\eta)\sum_{i=1}^{p+1} (dx^i)^2
\ee
where the scale factor is given by $S(\tilde\eta) \sim (\tilde\eta -\tilde\eta_0)^{1-\frac{p}{p+1}}$. This clearly shows that the
universes expand with deceleration for all $p$. For $p=2$, we get $S(\tilde\eta) \sim (\tilde\eta - \tilde\eta_0)^{\frac{1}{3}}$, 
the result that was obtained in \cite{Townsend:2003fx}.

Thus we have seen how starting from isotropic SD$p$ brane solutions of type II string theories, 
we get $(p+1)+1$ dimensional de Sitter spaces upto a conformal
factor by compactifying on $(8-p)$ dimensional hyperbolic spaces. This brings out the connection between space-like branes and the
de Sitter space which might be helpful in understanding dS/CFT correspondence in the same spirit as AdS/CFT correspondence. From the
metrics in this case we find that the space-times undergo decelerating expansion for all $p$, but only in  particular conformal
frames we get de Sitter spaces, i.e., eternal accelerations.  

\vspace{.5cm}

\noindent{\it 5. Conclusion} : To conclude, in this paper we have studied the various cosmological scenarios that are obtained from
the isotropic space-like D$p$ brane solutions of type II string theories by compactifications on $(8-p)$ dimensional hyperbolic
spaces and also found the connection between SD$p$ branes and $(p+1)+1$ dimensional de Sitter spaces. The SD$p$ brane solutions are 
characterized by three independent parameters $\tau_0$, $\theta$ and $\delta_0$. $\tau_0$ sets
a time scale in the theory, $\theta$ is related to the RR charge associated with SD$p$ branes and
$\delta_0$ is associated with the dilaton in the sense that when $p=3$, the dilaton is trivial for $\delta_0=0$ much like
time-like D3 branes. $\tau_0$ gives a time scale because when $\tau \gg \tau_0$, the SD$p$ brane solutions reduce to flat
spaces and in that sense these solutions are asymptotically ($\tau \to \infty$) flat. At large time or in the far future we found that
the external space-times undergo decelerating expansions where the scale factors behave like $S(\eta) \sim (\eta-\eta_0)^{1-\frac{p}{8}}$,
for all values of $p$ from 1 to 6. On the other hand, when $\tau \sim \tau_0$, the SD$p$ branes upon compactifications by
hyperbolic spaces give external space-times which in suitable coordinate can be recast into flat FLRW forms. Here we kept the parameter
$\delta_0$ to be arbitrary and found that $(p+1)+1$ dimensional external spaces undergo accelerating expansions for all $p$.
We have studied various cases numerically; because of the 
complicated nature of the solutions, it is not possible to study them analytically. We have plotted the expansion parameter $m(\tau)$ and 
the rate of expansion parameter $n(\tau)$ defined
in the text, for various values of $p$ to show the cosmologies in various dimensions. We found that for all $p$ lying between
1 to 6, there is a region
where $n(\tau)$ becomes positive for certain finite interval of time indicating that universes undergo a transient
phase of accelerating expansion. We have also plotted $n(\tau)$ when we vary 
the three parameters $\theta$, $\delta_0$ and $\tau_0$ while keeping the other parameters fixed in Figures 3, 4, and 5 respectively.
These show how the acceleration changes as we vary the parameters. Finally, we have shown that at early time, i.e., for $\tau \ll \tau_0$,
the $(p+1)+1$ dimensional external spaces can be cast into the form of de Sitter metrics upto a conformal transformation for all
values of $p$. Here we have fixed the parameter $\delta_0$ for calculational simplicity. This brings out the connection between the
SD$p$ branes and the de Sitter space which was the original motivation for constructing the space-like branes, and might be useful
in understanding dS/CFT correspondence in the same spirit as AdS/CFT correspondence. We mentioned that the cosmologies here again
are decelerating, but they give eternal accelerations only in a special conformal frame. 

\vspace{.5cm}

\noindent{\it Acknowledgements} : We would like to thank Koushik Dutta for helpful discussions.

\end{document}